\documentclass[aps,pra,epsfig,twocolumn,showpacs,superscriptaddress]{revtex4}
\usepackage{graphicx}
\usepackage{dcolumn}
\usepackage{bm}

\newcommand {\RR} {\mathrm{I\!R\!}}

\begin{document}

\title{Winning strategies for pseudo-telepathy games using single non-local box}
\author{Samir Kunkri}
\email{skunri_r@isical.ac.in}
\affiliation{Physics and Applied Mathematics Unit,
Indian Statistical Institute, 203 B.T. Road, Kolkata 700 108,
India}
\author{Guruprasad Kar}
\email{gkar@isical.ac.in}
\affiliation{Physics and Applied Mathematics Unit,
Indian Statistical Institute, 203 B.T. Road, Kolkata 700 108,
India}
\author{Sibasish Ghosh}
\email{sibasish@cs.york.ac.uk}
\affiliation{Department of Computer Science , The
University of York, Heslington, York, YO10  5DD, United
Kingdom}
\author{Anirban Roy}
\email{aroy@ictp.it}
\affiliation{The Abdus Salam International Centre for Theoretical Physics,
Strada Costiera 11, Trieste, Italy, 34014}

\date{\today}



\begin{abstract}
Using a single NL-box, a winning strategy is given for the
impossible colouring pseudo-telepathy game for the set of vectors
having Kochen-Specker property in four dimension. A sufficient
condition to have a winning strategy for the impossible colouring
pseudo-telepathy game for general $d$-dimension, with single use of
NL-box, is then described. It is also shown that the magic square
pseudo-telepathy game of any size can be won by using just two ebits
of entanglement -- for quantum strategy, and by a single NL-box --
for non-local strategy.
\end{abstract}


\maketitle

\section{Introduction}
By performing measurement on an entangled quantum system two
separate observer can obtain correlations that are nonlocal, in
the sense that no local hidden variable (LHV) model can reproduce
it. This was first proved by Bell in 1964 in terms of Bell
inequality \cite{bell}. Later on Clauser, Horne, Shimony and Holt
gave an experimental proposition of Bell's inequality which is
known as as CHSH inequality \cite{chsh}. According to CHSH
inequality all local hidden variable model must satisfy:
$$|\langle A_1B_1\rangle + \langle A_1B_2 \rangle + \langle A_2B_1
\rangle - \langle A_2B_2 \rangle|\le 2$$
 where $A_1$, $A_2$ are observables of a spin-half
particle in the possession of Alice and $B_1$, $B_2$ are
observables of a spin-half particle in the possession of Bob. But
local measurement carried out on entangled quantum system can
reach the value $2\sqrt{2}$. Cirelson's showed \cite{cirelson}
that this is the maximum value attainable by local measurement on
entangled quantum system although the maximum nonlocal value of
CHSH inequality can reach is $4$.

Popescu and Rohrlich \cite{PR} asked a very interesting question:
why quantum mechanics is not maximally non-local? Is there any
stronger correlation than the quantum mechanical ones that do not
allow signalling like quantum correlation? They have introduced a
hypothetical non-local box (NL box for short) that does not allow
signalling, yet violates CHSH inequality maximally. This NL-box
has two input bits $x$ and $y$, and yields two output bits $a$ and
$b$. The bits $x$ and $a$ are in Alice's hand, while $y$ and $b$
are in Bob's hand. The box is such that $a$ and $b$ are correlated
according to simple relation:
$$x.y = a \oplus b,$$
where $\oplus$ is addition modulo $2$. Afterwards many works have
been done to characterize the NL-box in order to yield insights
about the non-locality aspects of quantum mechanics
[\onlinecite{massar} -  \onlinecite{brassard1}].

Quantum pseudo-telepathy game \cite{brassard} provides an intuitive
way to understand quantum non-locality. Quantum pseudo-telepathy
game is something which can not be won in the classical world
without communication but can be won in the quantum world using
entangled state without any use of classical communication. Thus,
for an observer (ignorant about any sort of non-locality), the
reason for winning of the game by the players would imply some {\it
apriori} `telepathic' connection between the players. Nevertheless,
that sort of connection is impossible. Formally, according to ref.
\cite{brassard}, a two-party \cite{note1} pseudo-telepathy game is
given by a six-tuple $(X_A, X_B, Y_A, Y_B, P, W)$ where $X_A$ and
$X_B$ are the input sets of parties Alice and Bob respectively,
$Y_A$ and $Y_B$ are their respective output sets, $P$ ($\subseteq
X_A \times X_B$) is the set of all promises, and $W$ ($\subseteq X_A
\times X_B \times Y_A \times Y_B$) is the winning condition. Thus
$W$ is a relation between inputs and outputs that has to be
satisfied by Alice and Bob whenever the promise is fulfilled. Once
the respective inputs are supplied to Alice and Bob, they will no
longer be allowed to communicate until the game is over. In each
round of the game, Alice and Bob are supplied with the inputs $x \in
X_A$ and $y \in X_B$ respectively. Their task is now to produce
outputs $a \in Y_A$ and $b \in Y_B$ respectively. They will win the
round if either $(x, y) \notin P$ or $(x, y, a, b) \in W$. They will
win the game if they go on winning round after round. They will have
a winning strategy for the game if they are mathematically certain
to win the game as long as they have not exhausted all the classical
information as well as quantum entanglement (if there is any) shared
at the beginning of the game. Note that some observer of the game
(other than Alice and Bob) can only have a statistical evidence
towards making the hypothesis that Alice and Bob indeed have a
winning strategy for the game, if Alice and Bob go on winning the
game round after round. Quantum pseudo-telepathy games are proofs of
non-locality. Moreover, they are stronger proofs than usual Bell
theorems as well as Bell theorems without inequalities
\cite{methot}.

To understand the features of the NL-box it is necessary to
understand its power in various quantum information processing
protocols already discovered. There are entangled states (both
bipartite as well as multi-partite), the measurement correlations of
which can be simulated by one or more than one NL-box \cite{cerf,
gisin, barrett}. But there are measurement correlations
corresponding to some muti-partite entangled states which can not be
simulated by NL-boxes. In this context it would be interesting to
know whether all the pseudo-telepathy games, proposed so far, can be
won with single use of the NL-box. Recently Broadbent and
M$\acute{{\rm e}}$thot \cite{broa} showed that  some of the
pseudo-telepathy games can be won with single use of the NL-box
where the quantum strategy requires more than a maximally entangled
pair of qubits to succeed. It remained unsolved whether impossible
colouring pseudo-telepathy game, constructed by using Kochen-Specker
theorem, can be won with a single use of the NL-box. The problem, in
general, will be extremely difficult as there are various sets of
vectors satisfying Kochen-Specker property for Hilbert space of
dimension three or more. On the other hand it is known that the
magic square pseudo-telepathy game of {\it size} three ({\it i.e.},
where the size of the magic square matrix $3$) can be won by using a
single NL-box \cite{broa}. Whether this game, for any general size,
can be won by a single NL-box is also an unresolved issue. There is
a quantum winning strategy for the magic square pseudo-telepathy
game of size three that uses two ebits of shared entanglement
between the parties \cite{brassard}. The corresponding situation for
general size is unknown. As the magic square game of even size (see
section \ref{sec:MS} for the definition of magic square game) does
always have a classical solution ({\it i.e.}, the players are
neither required to use classical communication, nor to share any
entanglement, nor any NL-box), so we need to consider here only
games each having odd size.

In this paper first we shall present a winning strategy of
impossible colouring pseudo-telepathy game for the set of 18 vectors
having Kochen- Specker property in four dimension with single use of
NL-box. Then we discuss some sufficient condition for the winning
strategy of impossible colouring pseudo-telepathy game for general
$d$-dimension with single use of NL-box. We shall show here that the
magic square pseudo-telepathy game of any odd size\thanks{Unlike its
literary sense, here `odd size' refers to odd integral size.} can
also be won by using a single NL-box. Moreover, we shall describe a
quantum winning strategy for this game (of any odd size) which
requires only two ebits of shared entanglement between the parties.

In section \ref{sec:KS}, we shall describe the Kochen-Specker
theorem in four dimension that uses eighteen vectors from $\RR^4$,
and its corresponding impossible colouring pseudo-telepathy game is
described in section \ref{sec:tele}. A winning strategy for this
game is described in section \ref{sec:NL} using only one NL-box. A
winning strategy for the impossible colouring pseudo-telepathy games
in $d$ dimension, each of which satisfies a suitable sufficient
condition, is described in section \ref{sec:dNL} where it uses a
single NL-box. The magic square problem for general odd dimension is
described in section \ref{sec:MS}, where it is then posed as a
pseudo-telepathy game. A non-local winning strategy for this
pseudo-telepathy game is described in section \ref{sec:dMS} by using
a single NL-box. For the sake of completeness, the quantum winning
strategy for the magic square pseudo-telepathy game of size three is
described briefly in section \ref{sec:strat} which uses two ebits of
shared entanglement between the players. This strategy is used in
section \ref{sec:winMS} to provide a quantum winning strategy for
the magic square pseudo-telepathy game any general odd size by using
again only two ebits of shared entanglement between the players.
Section \ref{sec:con} draws the conclusion.

\section{\label{sec:KS}Kochen-Specker Theorem}

There exits an explicit, finite set of vectors in Hilbert space
with dimension $d \ge 3$, that can not be assigned values  $\{0,
1\}$ such that both of the conditions holds:
\begin{enumerate}
\item For every complete set of orthogonal basis vectors, only one
vector will get value $1$.
 \item Value assignment of the vectors will be non-contextual.
\end{enumerate}
We call such set of vectors a set with Kochen-Specker property.
\\

{\bf Example:}\\

 The following set of 18 (unnormalized) vectors in $\RR^4$
appearing in 9 sets of orthogonal basis  has Kochen- Specker
property \cite{cabello}. If on the contrary, one assumes  that
this set satisfy  both the conditions (1) and (2), one gets the
following equations.

$$V(0, 0, 0, 1) + V(0, 0, 1, 0)+ V(1, 1, 0, 0) + V(1, -1, 0, 0)= 1$$
$$V(0, 0, 0, 1) + V(0, 1, 0, 0)+ V(1, 0, 1, 0)+ V(1, 0, -1, 0) = 1$$
$$V(1, -1, 1, -1) + V(1, -1, -1, 1) + V(1, 1, 0, 0) + V(0, 0, 1, 1) = 1$$
$$V(1, -1, 1, -1) + V(1, 1, 1, 1) + V(1, 0, -1, 0) + V(0, 1, 0, -1) = 1$$
$$V(0, 0, 1, 0) + V(0, 1, 0, 0) + V(1, 0, 0, 1) + V(1, 0, 0, -1) = 1$$
$$V(1, -1, -1, 1) + V(1, 1, 1, 1) + V(1, 0, 0, -1) + V(0, 1, -1, 0) = 1$$
$$V(1, 1, -1, 1) + V(1, 1, 1, -1) + V(1, -1, 0, 0) + V(0, 0, 1, 1) = 1$$
$$V(1, 1, -1, 1) + V(-1, 1, 1, 1) + V(1, 0, 1, 0) + V(0, 1, 0, -1) = 1$$
$$V(1, 1, 1, -1) + V(-1, 1, 1, 1) + V(1, 0, 0, 1) + V(0, 1, -1, 0) = 1$$
Here $V(0, 0, 0, 1), \ldots, V(0, 1, -1, 0)$ denote the values taken
from the set $\{0, 1\}$ and are assigned to the respective vectors
$(0, 0, 0, 1), \ldots, (0, 1, -1, 0)$ (of $\RR^4$). If one add these
nine equations, the left hand side will be even as every vector has
appeared twice and their value can be $1$ or $0$, while the right
hand side is obviously odd. It proves that  one can not assign
values to  all vectors satisfying both the conditions.

\section{\label{sec:tele}Impossible colouring pseudo-telepathy game in 4-dimension}

We now turn this Kochen-Specker theorem in to a pseudo-telepathy
game as suggested by Brassard et al. \cite{brassard}. Consider the
nine complete orthogonal bases of real vectors in four dimension,
described in the above-mentioned example. Denote them by $S^1, S^2,
\ldots, S^9$, where each $S^J$ contains the following four pairwise
orthogonal vectors $u_1^J$, $u_2^J$, $u_3^J$, and $u_4^J$ where,
$u_1^1 = u_1^2 = (0, 0, 0, 1)$, $u_2^1 = u_1^5 = (0, 0, 1, 0)$, etc.
Two players, say, Alice and Bob, are far apart from each other such
that Alice is supplied with, at random, any one ($S^k$, say) of the
nine bases mentioned above, while Bob is supplied with, at random, a
vector ($u_m^l$, say) from the above-mentioned eighteen vectors. The
promise of the game is that $u_m^l$ must be a member of $S^k$. This
round of the game will be won by Alice and Bob if the following
conditions are satisfied:

Alice will have to assign value (0 or 1) to her four vectors
$u_1^k$, $u_2^k$, $u_3^k$, $u_4^k$ and Bob also will have to assign
value (0 or 1) to his single vector $u_m^k$ in such a way that

\begin{enumerate}
\item{Exactly one of Alice's four vectors should receive the value
1.}
\item{Alice and Bob have to assign same value to their single
common vector $u_m^k$.}
\end{enumerate}

with the condition that they will not be allowed to have any
classical communication after the game starts and until the game is
over. Thus they will win the game if they go on winning it for every
round of the game. Interestingly, Brassard et al. \cite{brassard}
presented a quantum winning strategy for a general impossible
colouring game in $d$ dimension using ${\rm log}_2 d$ ebits of
shared entanglement between Alice and Bob.

\section{\label{sec:NL}Wining strategy using a single NL-Box}
Now we shall present a strategy to win this game by using a single
NL-box. If one tries to satisfy all the above nine equations by
assigning non-contextual values to the maximum possible no. of
vectors, then one would see that seventeen vectors can be assigned
non-contextual values and value assignment for the remaining one
vector has to be contextual, {\it i.e.}, one vector out of eighteen
has to take value 1 when it occurs in one basis and 0 when it occurs
in another basis \cite{kar}.

Let us now consider a contextual value assignment to the vector $(0,
1, -1, 0)$, which appeared in the above-mentioned nine equations
twice -- once in the basis $S^6$ and once in $S^9$. We call the
following (contextual) value assignment strategy for this vector
(together with the remaining seventeen vectors) as $A0$: The vector
$(0, 1, -1, 0)$ takes value 1 when it occurs in $S^6$ and 0 when in
$S^9$; and the values assigned to the remaining seventeen vectors
are done non-contextually. Similarly we consider another contextual
value assignment (call it as $A1$) where the vector $(0, 1, -1, 0)$
take value 1 when it appears in $S^9$ and 0 when in $S^6$, value
assignments for the remaining vectors being non-contextual. Let $B0$
be the strategy where the eighteen vectors $u_1^1 = u_1^2 = (0, 0,
0, 1)$, $u_2^1 = u_1^5 = (0, 0, 1, 0)$, $u_3^1 = u_3^3 = (1, 1, 0,
0)$, $u_4^1 = u_3^7 = (1, -1, 0, 0)$, $u_2^2 = u_2^5 = (0, 1, 0,
0)$, $u_3^2 = u_3^8 = (1, 0, 1, 0)$, $u_4^2 = u_3^4 = (1, 0, -1,
0)$, $u_1^3 = u_1^4 = (1, -1, 1, -1)$, $u_2^3 = u_1^6 = (1, -1, -1,
1)$, $u_4^3 = u_4^7 = (0, 0, 1, 1)$, $u_2^4 = u_2^6 = (1, 1, 1, 1)$,
$u_4^4 = u_4^8 = (0, 1, 0, -1)$, $u_3^5 = u_3^9 = (1, 0, 0, 1)$,
$u_4^5 = u_3^6 = (1, 0, 0, -1)$, $u_4^6 = u_4^9 = (0, 1, -1, 0)$,
$u_1^7 = u_1^8 = (1, 1, -1, 1)$, $u_2^7 = u_1^9 = (1, 1, 1, -1)$,
$u_2^8 = u_2^9 = (-1, 1, 1, 1)$, appeared above, are assigned the
values 1, 0, 0, 0, 0, 0, 0, 1, 0, 0, 0, 0, 1, 0, 1, 1, 0, 0
respectively. Similarly, let $B1$ be the strategy where these
eighteen vectors (in the same order as above) are assigned the
values 1, 0, 0, 0, 0, 0, 0, 1, 0, 0, 0, 0, 0, 1, 0, 1, 0, 0
respectively. See the tables below for concise description of $A0$,
$A1$, $B0$ and $B1$.

Let each of Alice and Bob adopts two strategies: $A0$ and $A1$ for
Alice and $B0$ and $B1$ for Bob. One can now check that if, in any
round of the game, Alice adopts the strategy $A0$ and Bob adopts the
strategy $B0$, they will win that round of the game for all the
cases except when Alice is supplied with the basis $S^9$ and Bob is
supplied with the vector $u_4^6 = u_4^9 = (0~ 1~ -1~ 0)$. Same will
hold good if, instead, Alice adopts the strategy $A1$ while Bob
adopts $B1$. On the other hand both pairs of strategies $(A0, B1)$
and $(A1, B0)$ will give the winning condition of the game when
Alice is supplied with the basis $S^9$ and Bob has given the vector
$u_4^6 = u_4^9 = (0~ 1~ -1~ 0)$. The strategies $A0, B0, A1, B1$ are
given in the following tabular form:

\vspace{0.3cm}

\begin {tabular}{|l||c|}
\hline
\multicolumn{2}{|c|}{$A0$}\\
\cline{1-2}
\hline set  & value  \\
\hline $S^1$ & 1  0  0 0\\
\hline $S^2$ & 1  0  0 0\\
\hline $S^3$ & 1  0  0 0\\
\hline $S^4$ & 1  0  0 0\\
\hline $S^5$ & 0  0  1 0\\
\hline $S^6$ & 0  0  0 1\\
\hline $S^7$ & 1  0  0 0\\
\hline $S^8$ & 1  0  0 0\\
\hline $S^9$ & 0  0  1 0\\
\hline
\end{tabular}
\begin {tabular}{|l||c|}
\hline
\multicolumn{2}{|c|}{$B0$}\\
\cline{1-2}
\hline set  & value \\
\hline $S^1$ & 1  0  0 0\\
\hline $S^2$ & 1  0  0 0\\
\hline $S^3$ & 1  0  0 0\\
\hline $S^4$ & 1  0  0 0\\
\hline $S^5$ & 0  0  1 0\\
\hline $S^6$ & 0  0  0 1\\
\hline $S^7$ & 1  0  0 0\\
\hline $S^8$ & 1  0  0 0\\
\hline $S^9$ & 0  0  1 1\\
\hline
\end{tabular}
\begin {tabular}{|l||c|}
\hline
\multicolumn{2}{|c|}{$A1$}\\
\cline{1-2}
\hline set  & value  \\
\hline $S^1$ & 1  0  0 0\\
\hline $S^2$ & 1  0  0 0\\
\hline $S^3$ & 1  0  0 0\\
\hline $S^4$ & 1  0  0 0\\
\hline $S^5$ & 0  0  0 1\\
\hline $S^6$ & 0  0  1 0\\
\hline $S^7$ & 1  0  0 0\\
\hline $S^8$ & 1  0  0 0\\
\hline $S^9$ & 0  0  0 1\\
\hline
\end{tabular}
\begin {tabular}{|l||c|}
\hline
\multicolumn{2}{|c|}{$B1$}\\
\cline{1-2}
\hline set  & value \\
\hline $S^1$ & 1  0  0 0\\
\hline $S^2$ & 1  0  0 0\\
\hline $S^3$ & 1  0  0 0\\
\hline $S^4$ & 1  0  0 0\\
\hline $S^5$ & 0  0  0 1\\
\hline $S^6$ & 0  0  1 0\\
\hline $S^7$ & 1  0  0 0\\
\hline $S^8$ & 1  0  0 0\\
\hline $S^9$ & 0  0  0 0\\
\hline
\end{tabular}

\vspace{0.3cm}

Let us now assume that Alice and Bob share an NL-box. Alice and Bob
use this NL-box to choose their strategies among those alternatives.
The protocol is as follows: if Alice is supplied with one of the
first eight bases, {\it i.e.}, $S^1$, $S^2$, $\ldots$, $S^8$, then
she will provide $0$ as input to the NL-box, otherwise she will
choose $1$ as input. On the other hand if Bob is given the vector
$(0~ 1~ -1~ 0)$ he will provide $1$ as input, otherwise he will
choose $0$ as input to the NL-box. They will now select their
strategies according to the outputs of the NL-box, {\it i.e.}, if
Alice gets $0$ ($1$) as output, she will use the strategy $A0$
($A1$). Similarly if Bob gets $0$ ($1$) as output of the NL-box,
then he will assign value to the vector given to him according to
the strategy $B0$ ($B1$).

When Alice is told to assign values to the vectors of one of the
bases $S^1$, $S^2$, $\ldots$, $S^8$ and Bob to any vector from that
basis, the output of the NL-box will be either $0, 0$ or $1, 1$ to
Alice and Bob respectively. Accordingly they will adopt either
strategy $(A0, B0)$ or $(A1, B1)$. It is easy to verify from the
table that, in each case under this scenario, Alice and Bob will
assign same value to the vector given to Bob. When Alice's job is to
assign values to the vectors in the basis $S^9$ and Bob to any
vector except $(0, 1, -1, 0)$, the strategy will again be either
$(A0, B0)$ or $(A1, B1)$ and again it will work, as described above.
Only when Alice is asked to assign values to the vectors from the
basis $S^9$ and Bob for vector $(0, 1, -1, 0)$, both will put the
input $1$ in the NL-box and get either $0, 1$ or $1, 0$ as their
respective outputs. Here the strategy will be either $(A0, B1)$ or
$(A1, B0)$. The vector $(0, 1, -1, 0)$ has same value for both the
players. So the above-mentioned method produces a winning strategy
for the impossible colouring pseudo-telepathy game.

\section{\label{sec:dNL}Winning strategy for d-dimension using NL-box}
Constructions of sets of vectors in general $d$ dimensions (where $d
\ge 3$), having Kochen-Specker property, have been done separately
by using geometric method \cite{gill} and also by extending a
construction in dimension $d$ to dimension $d + 1$ \cite{peres}.
Using each of these constructions, the above-mentioned impossible
colouring pseudo-telepathy game can be generalized for any set of
 vectors having Kochen-Specker property for any dimension $d$
for $d \ge 3$. Brassard et al. \cite{brassard} have shown that if
all the vectors are real, then there is always a quantum strategy to
win this game, where Alice and Bob will have to share a maximally
entangled state in $d \otimes d$ of the form
$$|\phi_{AB}\rangle = \frac{1}{\sqrt{d}}~\sum_{i=1}^d |i\rangle_A \otimes
|i\rangle_B$$

Let in $d$-dimension, there are $n$ number of vectors with which $r$
number of orthogonal basis sets are formed with Kochen-Specker
property. Now we give some sufficient condition on such sets
for winning the pseudo-telepathy game constructed from these sets :\\
Let one start assigning values to the vectors appearing in the sets,
ordered arbitrarily, in a non-contextual way to satisfy both the
conditions of Kochen-Specker theorem (generalized to $d$ dimension).
After a certain steps, one finds that this non-contextual value
assignments do not work for the remaining $k$ (say) number of sets.
Let us now assume that the following condition holds good:

{\noindent {\it Sufficient condition}: Let $m$($\le k $) be the
number of different vectors to appear in those $k$ sets such that no
two or more than two of these $m$ vectors appear in any one of these
$k$ sets.}

We now consider a value assignment strategy (call it as $A0$) to the
above-mentioned $n$ vectors in $d$ dimension in such a way that up
to the first $(r - k)$ orthogonal bases, we non-contextually assign
$\{0, 1\}$-values to all the $d(r - k)$ vectors, that appeared in
these $(r - k)$ bases, maintaining both the conditions of
Kochen-Specker theorem, while (i) each of the above-mentioned $m$
vectors have to be assigned values (contextually) different from
values already assigned to them when they appeared in first $(r -
k)$ sets. By reversing the values of these $m$ vectors in the last
$k$ sets, one can satisfy the condition (i) with non-contextual
value assignment of the remaining $(n - m)$ vectors. We call this
later strategy as $A1$. We also consider strategies $B0$ and $B1$:
$B0$ is the strategy where the $\{0, 1\}$-value assignment to each
of the above-mentioned $n$ vectors will be same as those used in the
strategy $A0$ except for the above-mentioned $m$ vectors, for each
of which, the value assignment will be same as in the strategy $A1$
for its ({\it i.e.} $A1$'s) assignment of values to these $m$
vectors appeared in the last $k$ sets (as described above).
Similarly $B1$ is the strategy where the $\{0, 1\}$-value assignment
to each of the above-mentioned $n$ vectors will be same as those
used in the strategy $A1$ except for the above-mentioned $m$
vectors, for each of which, the value assignment will be same as in
the strategy $A0$ for its ({\it i.e.}, $A0$'s) assignment of values
to these $m$ vectors appeared in the last $k$ sets.

One can now find a strategy to win the game with a single NL- box.
The protocol works as follows:

Alice can use either of the two strategies $A0$ and $A1$ . Similarly
Bob can use either of the two strategies $B0$ and $B1$. Let Alice
and Bob are sharing a NL-box. They have fixed their protocol in this
way: when Alice will get any one of the set given from those $k$
sets, then she will give $1$ as input to the NL-box, otherwise she
will input $0$. Similarly when Bob will get any one of those $m$
vectors whose value has to be contextual, then he will give $1$ as
input to the NL-box and she will input $0$ otherwise. They will use
their strategies according to the outputs of NL-box, as described
earlier. One can check that this is a winning strategy for Alice and
Bob. Interestingly, the examples of non-colourable 37 vectors in 26
sets in $\RR^3$ and 20 vectors in 11 sets in $\RR^4$ \cite{bub}
satisfy the sufficient condition given above.

Existence of a classical deterministic winning strategy for the
impossible colouring game ({\it i.e.}, a strategy which does not use
entanglement or NL-box or any communication but where Alice can
assign $\{0, 1\}$-values to the vectors of each supplied orthogonal
basis (appeared in the associated Kochen-Specker theorem) $x$ and
Bob can also assign $\{0, 1\}$-values to each supplied vector (which
is a member of $x$) such that both the conditions in the
Kochen-Specker theorem are satisfied) would amount to contradict the
Kochen-Specker theorem itself. Hence such a strategy can not exist.

\section{\label{sec:MS}Generalization of the magic square problem}
The magic square problem of size $n = 2d + 1$, where $d$ is any
positive integer, is given as follows:

{\it Provide an $n$ by $n$ square arrangement with entries from
the set $\{0, 1\}$ such that (i) the modulo $2$ sum of all the
elements in each row is $0$ and (ii) the modulo $2$ sum of all
elements in each column is $1$, when $n$ is an arbitrary odd
positive integer greater than $1$.}

\vspace{0.4cm} {\noindent {\bf Magic square problem as a
pseudo-telepathy game:}}

\vspace{0.3cm}
Let there be two players Alice and Bob. Alice is
supplied with an element $x^{(A)}$  from the set $\{1, 2, \ldots,
n\}$, and similarly, Bob is supplied with an element $x^{(B)}$ from
the set $\{1, 2, \ldots, n\}$. After receiving $x^{(A)}$, Alice will
have to produce a {\it row} vector $\left(y^{(A)}_{x^{(A)}1},
y^{(A)}_{x^{(A)}2}, \ldots, y^{(A)}_{x^{(A)}n}\right) \in \{0,
1\}^n$, and similarly, after receiving $x^{(B)}$, Bob will have to
produce a {\it column} vector $\left(y^{(B)}_{1x^{(B)}},
y^{(B)}_{2x^{(B)}}, \ldots, y^{(B)}_{nx^{(B)}}\right)^T$ (where
$\left(y^{(B)}_{1x^{(B)}}, y^{(B)}_{2x^{(B)}}, \ldots,
y^{(B)}_{nx^{(B)}}\right) \in \{0, 1\}^n$) such that the following
conditions are simultaneously satisfied:

(1) modulo 2 sum of $y^{(A)}_{x^{(A)}1}$, $y^{(A)}_{x^{(A)}2}$,
$\ldots$, $y^{(A)}_{x^{(A)}n}$  is equal to $0$,

(2) modulo 2 sum of $y^{(B)}_{1x^{(B)}}$, $y^{(B)}_{2x^{(B)}}$,
$\ldots$, $y^{(B)}_{nx^{(B)}}$ is equal to $1$, and

(3) $y^{(A)}_{x^{(A)}x^{(B)}} = y^{(B)}_{x^{(A)}x^{(B)}}$

for every possible choice of $x^{(A)} \in \{1, 2, \ldots, n\}$ and
for every possible choice of $x^{(B)} \in \{1, 2, \ldots, n\}$.

\vspace{0.3cm} Note that here the question is not to produce a
complete $n$ by $n$ square arrangement with $0$'s and $1$'s
satisfying (i) and (ii) (which is, in fact, impossible), rather to
provide a mathematical argument that would unquestionably establish
the potentiality of the strategy to win the pseudo-telepathy game
for every possible input pair. Classically there can't exist a
winning strategy for this pseudo-telepathy game: A deterministic
classical winning strategy will have to assign $\{0, 1\}$-values to
each of the $n^2$ entries of the magic square -- which is
impossible. And so, there can't be any probabilistic classical
winning strategy either \cite{note2}.

\vspace{0.4cm} It is to be noted here that there is {\it no}
restriction on the total number of the answers
$$\left(\left(y^{(A)}_{x^{(A)}1},
y^{(A)}_{x^{(A)}2}, \ldots, y^{(A)}_{x^{(A)}n}\right),
\left(y^{(B)}_{1x^{(B)}}, y^{(B)}_{2x^{(B)}}, \ldots,
y^{(B)}_{nx^{(B)}}\right)\right)$$
$$\in \{0, 1\}^n \times \{0,
1\}^n,$$ that Alice and Bob could give (if that is possible at
all), corresponding each question $\left(x^{(A)}, x^{(B)}\right)
\in \{1, 2, \ldots, n\} \times \{1, 2, \ldots, n\}$ in the magic
square problem.

\section{\label{sec:dMS}A non-local winning strategy of the magic square pseudo-telepathy game using a single NL-box}
Let us consider the following row vectors from $\{0, 1\}^n$:

$e_1 = (0, 1, 1, \ldots, 1)$, {\it i.e.}, all the elements,
starting from the second, are equal to $1$, while the first
element is equal to $0$,

$e_2 = (1, 0, 1, \ldots, 1)$, {\it i.e.}, all the elements,
starting from the third, are equal to $1$, while the first element
is equal to $1$ and the second element is equal to $0$,

$\large{.}$

$\large{.}$

$\large{.}$

$e_{n - 1} = (1, 1, \ldots, 1, 0, 1)$, {\it i.e.}, all the
elements up to $(n - 2)$th position are equal to $1$, while the
$(n - 1)$th element is equal to $0$ and the $n$th element is equal
to $1$,

$e_n = (1, 1, \ldots, 1, 0)$, {\it i.e.}, all the elements up to
$(n - 1)$th position are equal to $1$, while the $n$th element is
equal to $0$;

$f_1 = (0, 0, \ldots, 0)$, {\it i.e.}, all the elements are equal
to $0$,

$f_2 = (0, 0, \ldots, 0, 1, 1)$, {\it i.e.}, all the elements up
to $(n - 2)$th position are equal to $0$, while the $(n - 1)$th as
well as the $n$th elements are equal to $1$;

$g_1 = (0, 1, 1, \ldots, 1, 0)$, {\it i.e.}, all the elements,
starting from the second and up to the $(n - 1)$th, are equal to
$1$, while the first and the $n$th elements are equal to $0$,

$g_2 = (1, 0, 1, 1, \ldots, 1, 0)$, {\it i.e.}, all the elements,
starting from the third and up to the $(n - 1)$th, are equal to
$1$, while the first element is equal to $1$ and the second as
well as the $n$th elements equal to $0$,

$\large{.}$

$\large{.}$

$\large{.}$

$g_{n - 2} = (1, 1, \ldots, 1, 0, 1, 0)$, {\it i.e.}, all the
elements, starting from the first and up to the $(n - 3)$th, are
equal to $1$, while the $(n - 2)$th as well as the $n$th elements
are equal to $0$ and the $(n - 1)$th element is equal to $1$,

$g_{n - 1} = (1, 1, \ldots, 1, 0, 0)$, {\it i.e.}, all the
elements, starting from the first and up to the $(n - 2)$th, are
equal to $1$, while the $(n - 1)$th as well as the $n$th elements
are equal to $0$;

$h_1 = (1, 1, \ldots, 1)$, {\it i.e.}, all the elements are equal
to $1$.

Let us now consider the following two strategies (we call them as
A0 and A1) to be adopted by Alice:

{\noindent {\underline {Strategy A0:}} If Alice adopts the
strategy A0, then she will choose her row vectors according to the
following rule:

$\left(y^{(A)}_{11}, y^{(A)}_{12}, \ldots, y^{(A)}_{1n}\right) =
e_1$, $\left(y^{(A)}_{21}, y^{(A)}_{22}, \ldots,
y^{(A)}_{2n}\right) = e_2$, $\ldots$ $\ldots$, $\left(y^{(A)}_{(n
- 1)1}, y^{(A)}_{(n - 1)2}, \ldots, y^{(A)}_{(n - 1)n}\right) =
e_{n - 1}$, $\left(y^{(A)}_{n1}, y^{(A)}_{n2}, \ldots,
y^{(A)}_{nn}\right) = f_1$.}

\vspace{0.3cm} {\noindent {\underline {Strategy A1:}} If Alice
adopts the strategy A1, then she will choose her row vectors
according to the following rule:

$\left(y^{(A)}_{11}, y^{(A)}_{12}, \ldots, y^{(A)}_{1n}\right) =
e_1$, $\left(y^{(A)}_{21}, y^{(A)}_{22}, \ldots,
y^{(A)}_{2n}\right) = e_2$, $\ldots$ $\ldots$, $\left(y^{(A)}_{(n
- 2)1}, y^{(A)}_{(n - 2)2}, \ldots, y^{(A)}_{(n - 2)n}\right) =
e_{n - 2}$, $\left(y^{(A)}_{(n - 1)1}, y^{(A)}_{(n - 1)2}, \ldots,
y^{(A)}_{(n - 1)n}\right) = e_n$, $\left(y^{(A)}_{n1},
y^{(A)}_{n2}, \ldots, y^{(A)}_{nn}\right) = f_2$.}

\vspace{0.4cm} Similarly, let us consider the following two
strategies (we call them as B0 and B1) to be adopted by Bob:

{\noindent {\underline {Strategy B0:}} If Bob adopts the strategy
B0, then he will choose his column vectors according to the
following rule:

$\left(y^{(B)}_{11}, y^{(B)}_{21}, \ldots, y^{(B)}_{n1}\right)^T =
g_1^T$, $\left(y^{(B)}_{12}, y^{(B)}_{22}, \ldots,
y^{(B)}_{n2}\right)^T = g_2^T$, $\ldots$ $\ldots$,
$\left(y^{(B)}_{1(n - 1)}, y^{(B)}_{2(n - 1)}, \ldots,
y^{(B)}_{n(n - 1)}\right)^T = g_{n - 1}^T$, $\left(y^{(B)}_{1n},
y^{(B)}_{2n}, \ldots, y^{(B)}_{nn}\right)^T = h_1^T$.}

\vspace{0.3cm} {\noindent {\underline {Strategy B1:}} If Bob
adopts the strategy B1, then he will choose his column vectors
according to the following rule:

$\left(y^{(B)}_{11}, y^{(B)}_{21}, \ldots, y^{(B)}_{n1}\right)^T =
g_1^T$, $\left(y^{(B)}_{12}, y^{(B)}_{22}, \ldots,
y^{(B)}_{n2}\right)^T = g_2^T$, $\ldots$ $\ldots$,
$\left(y^{(B)}_{1(n - 2)}, y^{(B)}_{2(n - 2)}, \ldots,
y^{(B)}_{n(n - 2)}\right)^T = g_{n - 2}^T$, $\left(y^{(B)}_{1(n -
1)}, y^{(B)}_{2(n - 1)}, \ldots, y^{(B)}_{n(n - 1)}\right)^T =
h_1^T$, $\left(y^{(B)}_{1n}, y^{(B)}_{2n}, \ldots,
y^{(B)}_{nn}\right)^T = g_{n - 1}^T$.}

\vspace{0.4cm} Note that if Alice adopts the strategy A0 and Bob
adopts B0, all the three conditions (1), (2) and (3), mentioned
above, are simultaneously satisfied for all $\left(x^{(A)},
x^{(B)}\right) \in \{1, 2, \ldots, n\} \times \{1, 2, \ldots, n\}$
except when $\left(x^{(A)}, x^{(B)}\right) = (n, n)$ (In this
particular case, all these three conditions are not satisfied.).

If Alice adopts the strategy A1 and Bob adopts B1, all the three
conditions (1), (2) and (3), mentioned above, are simultaneously
satisfied for all $\left(x^{(A)}, x^{(B)}\right) \in \{1, 2,
\ldots, n\}\times \{1, 2, \ldots, n\}$ except when $\left(x^{(A)},
x^{(B)}\right) = (n, n)$ (In this particular case, all these three
conditions are not satisfied.).

If Alice adopts the strategy A1 and Bob adopts B0, all the three
conditions (1), (2) and (3), mentioned above, are simultaneously
satisfied when $\left(x^{(A)}, x^{(B)}\right) = (n, n)$.

If Alice adopts the strategy A0 and Bob adopts B1, all the three
conditions (1), (2) and (3), mentioned above, are simultaneously
satisfied when $\left(x^{(A)}, x^{(B)}\right) = (n, n)$.

Thus we see that Alice and Bob can win the magic square game with
{\it certainty} if they jointly adopt one of the two strategies
(A0, B0) or (A1, B1) whenever $\left(x^{(A)}, x^{(B)}\right)$ is
not equal to $(n, n)$, and if they jointly adopt one of the two
strategies (A0, B1) or (A1, B0) whenever $\left(x^{(A)},
x^{(B)}\right) = (n, n)$.

\vspace{0.3cm} Let us now assume that Alice and Bob share a single
NLB with Alice's input bit and Bob's input bit as $X^{(A)}$ and
$X^{(B)}$ respectively, and with Alice's output bit and Bob's
output bit as $Y^{(A)}$ and $Y^{(B)}$ respectively such that the
modulo $2$ sum of $Y^{(A)}$ and $Y^{(B)}$ is equal to the product
of $X^{(A)}$ and $X^{(B)}$. Now given the input $x^{(A)} \in (\{1,
2, \ldots, n\} - \{n\})$ to Alice, she will supply the input
$X^{(A)} = 0$ to the NLB, else she will supply the input $X^{(A)}
= 1$ to the NLB. Similarly, given the input $x^{(B)} \in (\{1, 2,
\ldots, n\} - \{n\})$ to Bob, he will supply the input $X^{(B)} =
0$ to the NLB, else he will supply the input $X^{(B)} = 1$ to the
NLB. For these inputs to the NLB, Alice will then adopt the
strategy A$Y^{(A)}$ and Bob will adopt the strategy B$Y^{(B)}$.

Thus we see that when $x^{(A)} \in (\{1, 2, \ldots, n\} - \{n\})$
and $x^{(B)} \in (\{1, 2, \ldots, n\} - \{n\})$, we have $X^{(A)}
= X^{(B)} = 0$, and hence $\left(Y^{(A)}, Y^{(B)}\right) = (0, 0)$
or $(1, 1)$. So Alice and Bob can either adopt the strategies A0
and B0 respectively, or they can (equally well) adopt the
strategies A1 and B1 respectively. And in both of these cases they
can successfully win the magic square game.

When $x^{(A)} \in (\{1, 2, \ldots, n\} - \{n\})$ and $x^{(B)} =
n$, we have $X^{(A)} = 0$, $X^{(B)} = 1$, and hence
$\left(Y^{(A)}, Y^{(B)}\right) = (0, 0)$ or $(1, 1)$. So Alice and
Bob can either adopt the strategies A0 and B0 respectively, or
they can (equally well) adopt the strategies A1 and B1
respectively. And in both of these cases they can successfully win
the magic square game.

When $x^{(A)} = n$ and $x^{(B)} \in (\{1, 2, \ldots, n\} -
\{n\})$, we have $X^{(A)} = 1$, $X^{(B)} = 0$, and hence
$\left(Y^{(A)}, Y^{(B)}\right) = (0, 0)$ or $(1, 1)$. So Alice and
Bob can either adopt the strategies A0 and B0 respectively, or
they can (equally well) adopt the strategies A1 and B1
respectively. And in both of these cases they can successfully win
the magic square game.

When $x^{(A)} = n$ and $x^{(B)} = n$, we have $X^{(A)} = X^{(B)} =
1$, and hence $\left(Y^{(A)}, Y^{(B)}\right) = (0, 1)$ or $(1,
0)$. So Alice and Bob can either adopt the strategies A0 and B1
respectively, or they can (equally well) adopt the strategies A1
and B0 respectively. And in both of these cases they can
successfully win the magic square game.

\section{\label{sec:strat}Winning strategy for the magic square game of size three with
entanglement} Brassard et al. \cite{brassard} have shown how to win
the magic square game for $n = 3$ with certainty by sharing a two
ebit entanglement between Alice and Bob. Let us describe that
protocol below.

Alice and Bob are two far apart parties. Alice possess two qubits
$a$ and $c$ while Bob possess another two qubits $b$ and $d$. Let
us now assume that Alice and Bob share the singlets
$\left|{\psi}^-\right\rangle_{ab} =
\frac{1}{\sqrt{2}}\left(|00\rangle_{ab} + |11\rangle_{ab}\right)$
and $\left|{\psi}^-\right\rangle_{cd} =
\frac{1}{\sqrt{2}}\left(|00\rangle_{cd} + |11\rangle_{cd}\right)$.
Thus we see that Alice and Bob share the following $2$-ebit state:

$|\Psi\rangle_{ac:bd} = \frac{1}{2}(|00\rangle_{ac} \otimes |11\rangle_{bd} - |01\rangle_{ac} \otimes |10\rangle_{bd}$
\begin{equation}
\label{eq1}
- |10\rangle_{ac} \otimes |
01\rangle_{bd} + |11\rangle_{ac} \otimes |00\rangle_{bd}).
\end{equation}

According to her input $1$, or $2$, or $3$, Alice will first apply
respectively the following $4 \times 4$ unitary operators on her
two qubits:

$U_1 = \frac{1}{\sqrt{2}}\left[
                                      \begin{array}{cccc}
                                        i & 0 & 0 & 1 \\
                                        0 & -i & 1 & 0 \\
                                        0 & i & 1 & 0 \\
                                        1 & 0 & 0 & i \\
                                      \end{array}
                                    \right],~~ U_2 =
                                    \frac{1}{2}\left[
                                               \begin{array}{cccc}
                                                                    i & 1 & 1 &
i \\
                                                                    -i & 1 & -1
& i \\
                                                                    i & 1 & -1
& -i \\
                                                                    -i & 1 & 1
& -i \\
                                                                  \end{array}
                                                                \right]$

\begin{equation}
\label{eq2} U_3 =
                 \frac{1}{2}\left[                                                            \begin{array}{cccc}
                                                                               -1
& -1 & -1 & 1 \\
                                                                               1
& 1 & -1 & 1 \\
                                                                               1
& -1 & 1 & 1 \\
                                                                               1
& -1 & -1 & -1 \\
                                                                             \end{array}
                                                                           \right].
\end{equation}

Similarly, according to his input $1$, or $2$, or $3$, Bob will
first apply respectively the following $4 \times 4$ unitary
operators on his two qubits:

$V_1 = \frac{1}{2}\left[
                                      \begin{array}{cccc}
                                        i & -i & 1 & 1 \\
                                        -i & -i & 1 & -1 \\
                                        1 & 1 & -i & i \\
                                        -i & i & 1 & 1 \\
                                      \end{array}
                                    \right],~~ V_2 =
                                    \frac{1}{2}\left[
                                               \begin{array}{cccc}
                                                                    -1 & i & 1
& i \\
                                                                    1 & i & 1 &
-i \\
                                                                    1 & -i & 1
& i \\
                                                                    -1 & -i & 1
& -i \\
                                                                  \end{array}
                                                                \right]$
\begin{equation}
\label{eq3}
                                                                V_3
                                                                =
                                                                \frac{1}{\sqrt{2}}\left[
                                                                             \begin{array}{cccc}
                                                                               1
& 0 & 0 & 1 \\
                                                                               -1
& 0 & 0 & 1 \\
                                                                               0
& 1 & 1 & 0 \\
                                                                               0
& 1 & -1 & 0 \\
                                                                             \end{array}
                                                                           \right].
\end{equation}

After this, both Alice as well as Bob measure their respective two
qubits in the computational basis $\{|00\rangle, |01\rangle, |
10\rangle, |11\rangle\}$. Let $\left|a_1 a_2\right\rangle_{ac}$ and
$\left|b_1 b_2\right\rangle{bd}$ be the outputs of Alice and Bob,
provided they would occur with some non-zero (joint) probability.
Then Alice will supply the row vector $\left(a_1, a_2, a_1 \oplus
a_2\right)$ and Bob will supply the column vector $\left(b_1, b_2,
b_1 \oplus b_2 \oplus 1\right)^T$ as their respective outputs of the
magic square pseudo-telepathy game.

One can check that, using this strategy, Alice and Bob will be able
to win the game with certainty.

\section{\label{sec:winMS}Winning strategy for the magic square game of any odd size with entanglement}
Next we consider the question of winning the magic square
pseudo-telepathy game of size $n = 2d + 1$ using entanglement where
$d$ is any positive integer greater than $1$. So $n - 3 = 2(d - 1)$
will always be an even positive integer, greater than or equal to
$2$.

Let us again assume that two far apart parties Alice and Bob share
the $2$-ebit state $|\Psi\rangle_{ac:bd}$, as given in (\ref{eq1}),
where qubits $a$, $c$ are in the possession of Alice and the qubits
$b$, $d$ are in the possession of Bob.

\vspace{0.3cm} {\noindent {{\bf 1.} If the input $\left(x^{(A)},
x^{(B)}\right)$ for Alice and Bob belongs to the subset $\{1, 2,
\ldots, n - 3\} \times \{1, 2, \ldots, n - 3\}$ of $\{1, 2,
\ldots, n\} \times \{1, 2, \ldots, n\}$, then they will provide
their outputs as follows:}}

Upon receiving the inputs $\left(x^{(A)}, x^{(B)}\right)$, Alice and
Bob first performs the unitary operators $U_1$ (given in equation
(\ref{eq2})) and $V_1$ (given in equation (\ref{eq3})) on their
respective two qubits. And then they perform measurements in the
computational basis on their respective two qubits. If $\left|a_1
a_2\right\rangle_{ac}$ and $\left|b_1 b_2\right\rangle_{bd}$ be the
outcomes of Alice and Bob respectively in the above measurements
(so, the probability of their joint occurrence must be positive),
then Alice will provide the output row vector according to the rule

$\left(y^{(A)}_{11}, y^{(A)}_{12}, \ldots, y^{(A)}_{1n}\right) = (1,
1, \ldots, 1, 0, 0, 0)$ ({\it i.e.}, all the elements, except the
last three, are equal to $1$), $\left(y^{(A)}_{21}, y^{(A)}_{22},
\ldots, y^{(A)}_{2n}\right) = (0, 0, \ldots, 0)$ ({\it i.e.}, all
the elements are equal to $0$),\\ $\left(y^{(A)}_{31}, y^{(A)}_{32},
\ldots, y^{(A)}_{3n}\right) = (0, 0, \ldots, 0)$, $\ldots$ $\ldots$,
$\left(y^{(A)}_{(n - 3)1}, y^{(A)}_{(n - 3)2}, \ldots, y^{(A)}_{(n -
3)n}\right) = (0, 0, \ldots, 0)$;

$\left(y^{(B)}_{11}, y^{(B)}_{21}, \ldots, y^{(B)}_{n1}\right)^T =
(1, 0, 0, \ldots, 0)^T$ ({\it i.e.}, all the elements, except the
first one, are equal to $0$), $\left(y^{(B)}_{12}, y^{(B)}_{22},
\ldots, y^{(B)}_{n2}\right)^T = (1, 0, 0, \ldots, 0)^T$,
$\left(y^{(B)}_{13}, y^{(B)}_{23}, \ldots, y^{(B)}_{n3}\right)^T =
(1, 0, 0, \ldots, 0)^T$, $\ldots$ $\ldots$, $\left(y^{(B)}_{1(n -
3)}, y^{(B)}_{2(n - 3)}, \ldots, y^{(B)}_{n(n - 3)}\right)^T = (1,
0, 0, \ldots, 0)^T$.

Thus we see that the outputs of Alice and Bob, in this case, does
not depend on the choice of the unitary operators and measurement
outcomes.

\vspace{0.3cm} {\noindent {{\bf 2.} If the input $\left(x^{(A)},
x^{(B)}\right)$ for Alice and Bob belongs to the subset $\{1, 2,
\ldots, n - 3\} \times \{n - 2, n - 1, n\}$, then they will
provide their outputs as follows:}}

Upon receiving the inputs $\left(x^{(A)}, x^{(B)}\right)$, Alice
and Bob first performs the unitary operators $U_1$ and $V_{x^{(B)}
- n + 3}$ on their respective two qubits. And then they perform
measurements in the computational basis on their respective two
qubits. If $\left|a_1 a_2\right\rangle_{ac}$ and $\left|b_1
b_2\right\rangle_{bd}$ be the outcomes of Alice and Bob
respectively in the above measurements (so, the probability of
their joint occurrence must be positive), then Alice will provide
the output row vector according to the rule

$\left(y^{(A)}_{11}, y^{(A)}_{12}, \ldots, y^{(A)}_{1n}\right) = (1,
1, \ldots, 1, 0, 0, 0)$, $\left(y^{(A)}_{21}, y^{(A)}_{22}, \ldots,
y^{(A)}_{2n}\right) = (0, 0, \ldots, 0)$,\\ $\left(y^{(A)}_{31},
y^{(A)}_{32}, \ldots, y^{(A)}_{3n}\right) = (0, 0, \ldots, 0)$,
$\ldots$ $\ldots$, $\left(y^{(A)}_{(n - 3)1}, y^{(A)}_{(n - 3)2},
\ldots, y^{(A)}_{(n - 3)n}\right) = (0, 0, \ldots, 0)$,

while Bob will provide the output column vector according to the
rule

$\left(y^{(B)}_{1x^{(B)}}, y^{(B)}_{2x^{(B)}}, \ldots,
y^{(B)}_{nx^{(B)}}\right)^T = (0, 0, \ldots, 0, b_1, b_2, b_1
\oplus b_2 \oplus 1)^T$ ({\it i.e.}, all the elements, except the
last three, are equal to $0$).

\vspace{0.3cm} {\noindent {{\bf 3.} If the input $\left(x^{(A)},
x^{(B)}\right)$ for Alice and Bob belongs to the subset $\{n - 2,
n - 1, n\} \times \{1, 2, \ldots, n - 3\}$, then they will provide
their outputs as follows:}}

Upon receiving the inputs $\left(x^{(A)}, x^{(B)}\right)$, Alice
and Bob first performs the unitary operators $U_{x^{(A)} - n + 3}$
and $V_1$ on their respective two qubits. And then they perform
measurements in the computational basis on their respective two
qubits. If $\left|a_1 a_2\right\rangle_{ac}$ and $\left|b_1
b_2\right\rangle_{bd}$ be the outcomes of Alice and Bob
respectively in the above measurements (so, the probability of
their joint occurrence must be positive), then Alice will provide
the output row vector according to the rule

$\left(y^{(A)}_{x^{(A)}1}, y^{(A)}_{x^{(B)}2}, \ldots,
y^{(A)}_{x^{(A)}n}\right) = (0, 0, \ldots, 0, a_1, a_2, a_1 \oplus
a_2)$ ({\it i.e.}, all the elements, except the last three, are
equal to $0$),

while Bob will provide the output column vector according to the
rule

$\left(y^{(B)}_{11}, y^{(B)}_{21}, \ldots, y^{(B)}_{n1}\right)^T =
(1, 0, 0, \ldots, 0)^T$, $\left(y^{(B)}_{12}, y^{(B)}_{22}, \ldots,
y^{(B)}_{n2}\right)^T = (1, 0, 0, \ldots, 0)^T$,\\
$\left(y^{(B)}_{13}, y^{(B)}_{23}, \ldots, y^{(B)}_{n3}\right)^T =
(1, 0, 0, \ldots, 0)^T$, $\ldots$ $\ldots$, $\left(y^{(B)}_{1(n -
3)}, y^{(B)}_{2(n - 3)}, \ldots, y^{(B)}_{n(n - 3)}\right)^T = (1,
0, 0, \ldots, 0)^T$.

\vspace{0.3cm} {\noindent {{\bf 4.} If the input $\left(x^{(A)},
x^{(B)}\right)$ for Alice and Bob belongs to the subset $\{n - 2,
n - 1, n\} \times \{n - 2, n - 1, n\}$, then they will provide
their outputs as follows:}}

Upon receiving the inputs $\left(x^{(A)}, x^{(B)}\right)$, Alice
and Bob first performs the unitary operators $U_{x^{(A)} - n + 3}$
and $V_{x^{(B)} - n + 3}$ on their respective two qubits. And then
they perform measurements in the computational basis on their
respective two qubits. If $\left|a_1 a_2\right\rangle_{ac}$ and
$\left|b_1 b_2\right\rangle_{bd}$ be the outcomes of Alice and Bob
respectively in the above measurements (so, the probability of
their joint occurrence must be positive), then Alice will provide
the output row vector according to the rule

$\left(y^{(A)}_{x^{(A)}1}, y^{(A)}_{x^{(B)}2}, \ldots,
y^{(A)}_{x^{(A)}n}\right) = (0, 0, \ldots, 0, a_1, a_2, a_1 \oplus
a_2)$,

while Bob will provide the output column vector according to the
rule

$\left(y^{(B)}_{1x^{(B)}}, y^{(B)}_{2x^{(B)}}, \ldots,
y^{(B)}_{nx^{(B)}}\right)^T = (0, 0, \ldots, 0, b_1, b_2, b_1
\oplus b_2 \oplus 1)^T$.

\vspace{0.4cm} Using this strategy, one can check (which is simple
but tedious) that Alice and Bob will be able to win the game with
certainty.

Note that instead of applying the unitary operator $U_1$, Alice
could also have applied any one (but fixed) of $U_2$, $U_3$ whenever
she receives her input $x^{(A)}$ from the subset $\{1, 2, \ldots, n
- 3\}$. Similarly, Bob also could have applied any one (but fixed)
of $V_1$, $V_2$, $V_3$ whenever he receives her input $x^{(B)}$ from
the subset $\{1, 2, \ldots, n - 3\}$.

\section{\label{sec:con}Conclusion}
One should note that in the discussion of the sufficient conditions
for having non-local winning strategy for the impossible colouring
pseudo-telepathy game in $d$ dimension, we have not mentioned that
vectors have to be real. That condition is necessary for quantum
protocol \cite{brassard} but has no relevance for protocol using
NL-box. This difference may be important to understand particular
features of quantum entanglement in the context of general non-local
theory with no signalling. As we go on increasing the size of the
inputs in a pseudo-telepathy game, we might expect of using more
resources both for quantum as well as for non-local winning
strategies. But our result proves it to be not true for the magic
square pseudo-telepathy game.

In order to characterize properties of the non-local correlation
associated to the NL-box, it is important to classify all possible
non-local correlations (including quantum one), each of which can be
simulated by one or more than one NL-box (without allowing any
communication). For example, the EPR correlation, for von Neumann
measurements, can be simulated by a single NL-box \cite{cerf}. The
quantum correlations, arising out from the quantum winning strategy
of the magic square pseudo-telepathy game of size three using two
EPR pairs, can be simulated by a single NL-box by using the method
for having a non-local winning strategy for the above game
\cite{broa}. It would be similarly interesting to see whether both
the quantum correlations -- one in impossible colouring
pseudo-telepathy game in $d$ dimension and another in general magic
square pseudo-telepathy game -- can be simulated by their
corresponding non-local winning strategies, each of which uses only
a single NL-box.

\section{Acknowledgement}
Authors would like to thank Prof. S. P. Pal and S. K. Choudhary for
useful discussions. Authors would also like to thank A. A.
M$\acute{{\rm e}}$thot for useful comments on the earlier version of
the paper and for pointing out the works described in references
\cite{methot} and \cite{barrett}. S. K. acknowledges the support by
CSIR, Government of India, New Delhi. Part of this work has been
done while SG and AR was visiting the Physics and Applied
Mathematics Unit of Indian Statistical Institute, Kolkata - 700 108,
India. This work of SG is funded in part by EPSRC grant GR/87406.


\begin{thebibliography}{99}
\bibitem{note1} We only consider here two-party quantum
pseudo-telepathy games.
\bibitem{note2} A probabilistic classical winning strategy for the
magic square pseudo-telepathy game requires implementation of some
strategy $(A1, B1)$ for Alice and Bob respectively with non-zero
probability $p_1$ (where this strategy can be implemented by Alice
and Bob by sharing some random variables and then tossing local
coins to get values of the local random variables, which, in turn,
will fix the strategy), such that the strategy can produce correct
outputs for all possible inputs with  probability  at least $p$
where $p > 0$. Note that $p$ has to be 0 here as any non-zero $p$
would amount to successfull implementation of a classical
deterministic winning strategy for the magic square pseudo-telepathy
game (see \cite{brassard} for details).
\end{thebibliography}
\end{document}